\documentclass[aps,prl,twocolumn,superscriptaddress,10pt,nobibnotes]{revtex4-2}

\usepackage{mathptmx}
\usepackage{graphicx}
\usepackage{amsmath}
\usepackage{amssymb}
\usepackage{braket}
\usepackage{booktabs}
\usepackage{xcolor}
\usepackage{colortbl}
\definecolor{lightblue}{rgb}{0.85,0.90,0.95}
\usepackage{tikz}
\usepackage{tcolorbox}
\tcbuselibrary{breakable}
\newtcolorbox{taskbox}{
  breakable,
  colback=gray!4,
  colframe=gray!50,
  boxrule=0.4pt,
  arc=1.5mm,
  left=6pt, right=6pt, top=5pt, bottom=5pt,
  fontupper=\small
}
\usepackage[hidelinks]{hyperref}
\hypersetup{pdftitle={Agent-Based Autonomous Calibration and Control of a Neutral-Atom Quantum Processor}}

\graphicspath{{./}{figs/}}

\begin{document}

\title{Agent-Based Autonomous Calibration and Control of a Neutral-Atom Quantum Processor}

\author{Yichao Guo}
\thanks{These authors contributed equally to this work.}
\affiliation{Beijing Academy of Quantum Information Sciences, Beijing 100193, China}
\affiliation{State Key Laboratory of Low Dimensional Quantum Physics, Tsinghua University, Beijing 100084, China}

\author{Cheng Ye}
\thanks{These authors contributed equally to this work.}
\affiliation{Institute of Theoretical Physics, Chinese Academy of Sciences, Beijing 100190, China}
\affiliation{QuEmerge, Beijing 100190, China}

\author{Ruiyi Wang}
\thanks{These authors contributed equally to this work.}
\affiliation{Beijing Academy of Quantum Information Sciences, Beijing 100193, China}

\author{Yuanjiang Tang}
\affiliation{State Key Laboratory of Low Dimensional Quantum Physics, Tsinghua University, Beijing 100084, China}

\author{Linyuan Jiang}
\affiliation{State Key Laboratory of Low Dimensional Quantum Physics, Tsinghua University, Beijing 100084, China}

\author{Shun Jiang}
\affiliation{Beijing Academy of Quantum Information Sciences, Beijing 100193, China}

\author{Yi-Quan Zou}
\email{zouyq@baqis.ac.cn}
\affiliation{Beijing Academy of Quantum Information Sciences, Beijing 100193, China}

\author{Li You}
\affiliation{Beijing Academy of Quantum Information Sciences, Beijing 100193, China}
\affiliation{State Key Laboratory of Low Dimensional Quantum Physics, Tsinghua University, Beijing 100084, China}
\affiliation{Frontier Science Center for Quantum Information, Beijing, China}
\affiliation{Hefei National Laboratory, Hefei, Anhui 230088, China}

\author{Pan Zhang}
\email{panzhang@itp.ac.cn}
\affiliation{Institute of Theoretical Physics, Chinese Academy of Sciences, Beijing 100190, China}
\affiliation{School of Physical Sciences, University of Chinese Academy of Sciences, Beijing 100049, China}
\affiliation{School of Fundamental Physics and Mathematical Sciences, Hangzhou Institute for Advanced Study, UCAS, Hangzhou 310024, China}

\begin{abstract}
Neutral-atom quantum processors, combining programmable atom arrays with Rydberg-mediated entanglement, have emerged as a leading platform for scalable quantum computing.
Maintaining their reliable operation requires recurrent calibration of interdependent optical and atomic control parameters across heterogeneous instruments.
The dynamic nature of experimental conditions calls for a flexible control framework that integrates instrument coordination, measurement interpretation, and calibration decisions.
Here we present an agent-orchestrated central control system that autonomously executes routine maintenance and connects the experiment control stack and all peripheral instruments behind a single agent interface.
The agent is powered by a general-purpose large language model (LLM) and operates through laboratory knowledge Skills distilled from experts.
The Skills are a modular library, in which the calibration sections are self-contained modules invoked on demand for recurrent maintenance tasks, while coordination modules record the operating conventions of each instrument family.
Safety boundaries for high-power laser operation are encoded in the library and enforced by deterministic checks outside the language model.
The agent executes the full spectrum of laboratory operations, from instrument verification, through clock-state preparation and single-qubit gate calibration, to Rydberg beam alignment and two-qubit gate optimization.
On a $^{87}$Rb Rydberg neutral-atom platform, the system executed calibration tasks over a 90-day deployment with more than 850 experimental engagements.
The results demonstrate autonomous maintenance under experimentalist supervision, extending agent-based laboratory operation to long-term experiment assistance.
\end{abstract}

\maketitle

\begin{figure*}[t]
\centering

\includegraphics[width=\textwidth]{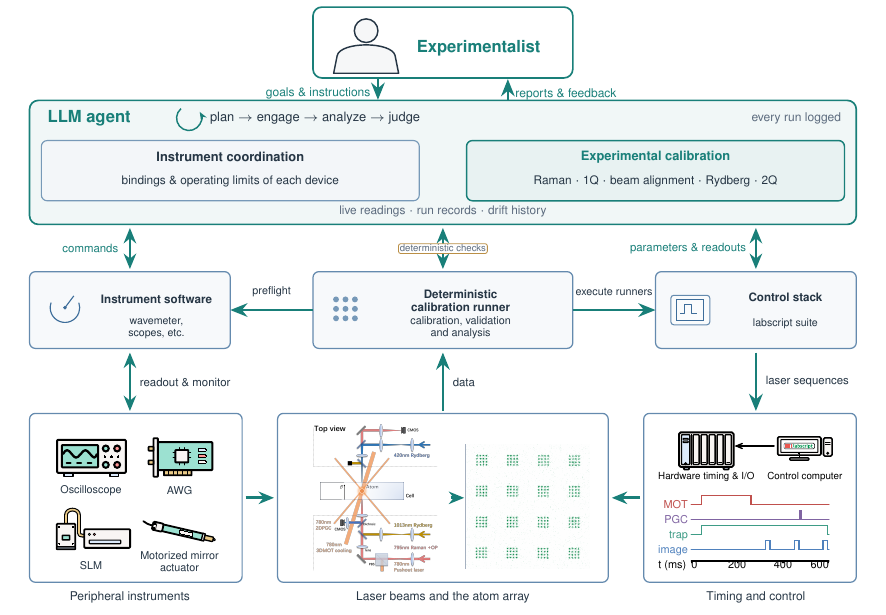}
\caption{\textbf{Agent-orchestrated central control of a neutral-atom platform.} The experimentalist provides goals and receives reports and requests for feedback. The LLM agent operates through a plan--engage--analyze--judge loop, drawing on instrument-coordination and experimental-calibration Skills. Instrument tools and deterministic preflight and safety checks mediate access to calibration runners and the labscript control stack, which execute parameter scans and laser sequences at the atom plane. Live readings and previous run records inform subsequent decisions, and every executed run is logged. Three interface layers connect the agent to the apparatus: the control software of the peripheral instruments, the deterministic calibration runner with its analysis tools, and the labscript control stack. The bottom row shows the corresponding hardware: the peripheral instruments; the atom plane, with the laser paths that reach the atoms and a fluorescence image of the rearranged 16$\times$16 array, arranged in 4$\times$4 blocks of 4$\times$4 sites, from a single shot in which all 256 sites were loaded, with the background subtracted for display (Supplemental Material); and the recorded timing of one experimental shot, comprising MOT loading, PGC, the tweezer trap and three camera exposures. Camera images return to the analysis layer, closing the loop.}
\label{fig:system}
\end{figure*}

Neutral-atom quantum processors have advanced rapidly in atom-array preparation, coherent control, and logical-qubit operation, establishing them as a leading platform for scalable quantum computing~\cite{endres2016,barredo2016,levine2019,evered2023,bluvstein2024,manetsch2025tweezer,chiu2025continuous,bluvstein2026architecture}.
These processors combine rearrangeable geometries, single-atom imaging, hyperfine-state control, and Rydberg-mediated interactions within a single experimental platform~\cite{browaeys2020,saffman2010,ebadi2021}.
As they advance toward larger and more complex systems, sustaining precise and reliable control requires recurrent calibration and coordination across interdependent experimental subsystems.
In our apparatus, optical control, waveform generation, and diagnostics expose separate interfaces across several computers, leaving the experimentalist to coordinate their operation.
Laser frequency, power, and beam alignment affect the atomic response jointly, while drift makes calibration recurrent~\cite{maller2015,kelly2016scalable,kelly2018dag,evered2023}.
Effective maintenance therefore requires connecting optical diagnostics to atomic measurements, assessing which calibrations remain valid, and selecting the next measurement from incomplete or noisy evidence.
This combination of instrument coordination, recurrent calibration, and evidence-dependent judgment motivates a control framework that couples deterministic experiment execution with intelligent workflow decisions.

Conventional control software automates individual measurements~\cite{labscript_suite}, and schedulers and rule engines organize established sequences~\cite{kelly2018dag,patra2021acting}.
Such software provides a reliable execution layer, but decisions above this layer, such as whether a calibration remains valid, whether a noisy result is acceptable, and when an anomaly should stop the workflow, still require experimental judgment.
A scalable maintenance system must connect these decisions to instrument actions without replacing the deterministic timing and execution layer.

Large language models (LLMs), coupled to tools and execution environments, offer a way to interpret instructions, combine experimental evidence with procedural knowledge, and translate decisions into instrument actions.
Such agents have advanced intelligent experimentation in chemical synthesis and materials discovery~\cite{boiko2023autonomous,szymanski2023autonomous,tom2024sdl}, the operation of scientific instruments and user facilities~\cite{uddin2026optics,kaiser2025accelerator,vriza2026,chen2026xray}, and quantum experiments~\cite{isogawa2026sensing,cao2024agents,li2026large}.
In particular, Vibe Calibration demonstrated autonomous calibration on a 112-qubit superconducting processor by a language agent orchestrating reusable Skills distilled from expert knowledge~\cite{lzy_vibe}.
Recent neutral-atom work has translated research protocols into experiments on cloud-accessible processors~\cite{dalyac2026agentic} and used an LLM with human feedback to guide acousto-optic tweezer-array optimization~\cite{sheng2026tweezerllm}.
These advances respectively address Skill-based device bring-up and calibration, protocol execution, and optimization of a specific optical subsystem.
What remains insufficiently demonstrated is sustained agent-based maintenance across a heterogeneous neutral-atom apparatus, combining cross-instrument coordination, validation through atomic measurements, and deterministic safety enforcement.

We present an agent-orchestrated central control system for a $^{87}$Rb Rydberg neutral-atom array that combines reusable Skills with deterministic experiment runners and safety checks.
The system unifies access to the control stack, peripheral instruments, and analysis tools, freeing experimentalists from routine cross-instrument coordination; they define objectives and retain authority over operations outside documented procedures.
During a 90-day deployment, the system completed more than 850 agent-executed experimental engagements, comprising complete maintenance rounds and individual calibration, verification, and readout tasks.
Representative experiments demonstrate clock-state preparation, single-qubit gate calibration, closed-loop Rydberg beam alignment using spatially resolved light-shift measurements, extraction of resonance offsets and Rabi frequencies from two-dimensional scans, and six-parameter CZ-gate optimization.
These results establish sustained and auditable agent-based maintenance across the local apparatus under explicit safety and authorization constraints.

To address the coordination requirements, the system in Fig.~\ref{fig:system} links an LLM agent, a Skill library, instrument tools, and the experimentalist.
The four stages shown in Fig.~\ref{fig:system}: plan, engage, analyze, and judge, correspond to selecting the relevant Skills, invoking instrument tools or deterministic runners, analyzing the returned measurements, and applying acceptance criteria or escalating an anomaly.
Skills encode procedures, constraints, and acceptance criteria; an optional notebook retains work-session context.
Instructions and feedback from the experimentalist, together with results and anomalies returned by the agent, form a bidirectional interaction.

The agent runs on a separate computer and drives the labscript control stack~\cite{labscript_suite} through its remote interface; most instruments are hardware-timed devices of this stack and act within the experimental sequence.
Instruments that operate outside the sequence, such as devices held at fixed settings or monitored continuously, were previously set by hand through their own control programs.
Bash/Python tools provide unified access to these instruments, experiment scripts, configuration files, and analysis utilities.
The Skill documentation includes reference descriptions of the nominal experimental sequence, relevant control parameters, and their physical meanings.
Experiment-control tools connect the agent to instrument operations and deterministic runners.
A runner selects the authorized sequence script, sets scan axes and timing parameters within the bounds specified by its Skill, launches the measurement through the control stack, and polls the resulting shot files for analysis.
This tool--runner--execution chain separates workflow decisions from low-level timing and hardware execution while allowing explicitly authorized development of new execution or analysis scripts.

Operational authority and safety constraints are enforced separately from the language model.
The experimentalist defines objectives, issues the instructions that initiate the engagements, authorizes operations, and resolves cases outside documented procedures; the agent returns measurement evidence and proposed next steps.
Scan ranges are specified in the Skills, and instrument-specific limits, such as bounds on actuator step sizes, are enforced by deterministic software in those control programs (Supplemental Material). 

When a required check fails or an action exceeds its authority, the agent halts and refers the decision to the experimentalist.

The Skill library organizes instrument coordination and experimental calibration into modular procedures with explicit operating conventions, execution constraints, and acceptance criteria.
Table~\ref{tab:skills} lists the deployed Skills of these two classes.
Each instrument-coordination Skill binds one instrument family and records its admissible command format, the semantics of each response, and the operations prohibited on that interface.
The device-status Skill interrogates the spatial light modulator, the waveform generators, and the oscilloscope used in the preflight checks described below.
This separation is designed to localize interface changes primarily to the instrument-coordination Skills when the experimental procedures and acceptance criteria remain applicable.

\begin{table}[t]
\caption{\textbf{Skill modules deployed on the neutral-atom platform.} Instrument coordination binds instruments and control interfaces; experimental calibration carries the procedures used in recurrent maintenance. The daily run log provides the audit trail described in the text.}
\label{tab:skills}
\small
\setlength{\tabcolsep}{4pt}
\begin{tabular}{ll}
\toprule
\textbf{Module} & \textbf{Function} \\
\midrule
\rowcolor{lightblue}
\multicolumn{2}{c}{\textit{Instrument coordination}} \\
\texttt{control-stack} & Script control; shot execution \\
\texttt{wavemeter-snapshot} & 8-channel laser readout \\
\texttt{device-status} & SLM/AWG/scope heartbeat \\
\texttt{mirror-control} & Rydberg beam stepping \\
\texttt{atom-array-detection} & Interface to occupancy analysis \\
\texttt{slm-rearrangement} & Target-geometry preparation \\
\texttt{scope-monitor} & Scope channel readout \\
\midrule
\rowcolor{lightblue}
\multicolumn{2}{c}{\textit{Experimental calibration}} \\
\texttt{raman-clock-prep} & Raman calibration; survival chain \\
\texttt{single-qubit-rb} & $\pi/2$ calibration; Clifford RB \\
\texttt{rydberg-beam-alignment} & Light-shift spatial profile \\
\texttt{rydberg-excitation} & $\Omega_0$, $\delta_0$ from 2D map \\
\texttt{two-qubit} & CZ scan; SSB benchmark \\
\bottomrule
\end{tabular}
\end{table}

Experimental calibration Skills are self-contained modules dedicated to individual calibration targets and are invoked on demand rather than traversed as a fixed decision tree.
These calibrations provide the recurrent foundation on which the platform's other experiments build.
Live readings and previous run records provide the evidence used by the agent to diagnose problems and to select or repeat calibration modules against the requirements of the task at hand; anomalies beyond its authority are escalated to the experimentalist.
Each experimental Skill records the valid timing requirements, writable parameters, returned-data structure, and acceptance criteria.
Its associated runner performs the authorized measurement, while the agent determines whether the result satisfies those criteria and what action follows.
The five experimental calibration Skills listed in Table~\ref{tab:skills} implement the maintenance workflow evaluated below.

The Skill library is revised on the basis of feedback from the experimentalists, and each revision is reviewed by an experimentalist before use.
For every engagement, the system appends a structured record (input parameters, analysis result, acceptance verdict, artifacts, and next action) to a daily log (see the Supplemental Material).
Together with the parameters archived in each shot file, these records make the executed workflow auditable: each engagement can be traced to its parameters and to the experimental evidence used for acceptance.

\begin{figure*}[t]
\centering

\includegraphics[width=\textwidth]{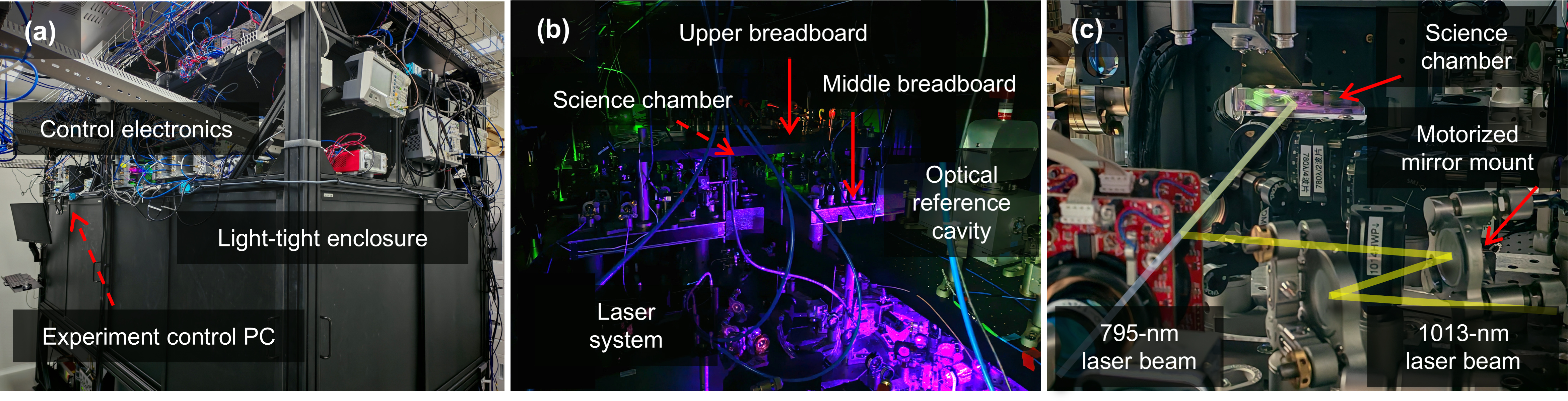}
\caption{\textbf{The neutral-atom apparatus on which the reported runs were recorded.} (a) Exterior view, with the control electronics above the light-tight enclosure and the experiment-control computer below. (b) The optical setup inside the enclosure, showing the laser system, the upper and middle breadboards around the science chamber, and the optical reference cavity. (c) Close-up of the science chamber with the 795-nm and 1013-nm beam paths, drawn as overlays, and one of the motorized mirror mounts stepped by the agent during beam alignment.}
\label{fig:apparatus}
\end{figure*}

\begin{figure*}[t]
\centering
\includegraphics[width=\textwidth]{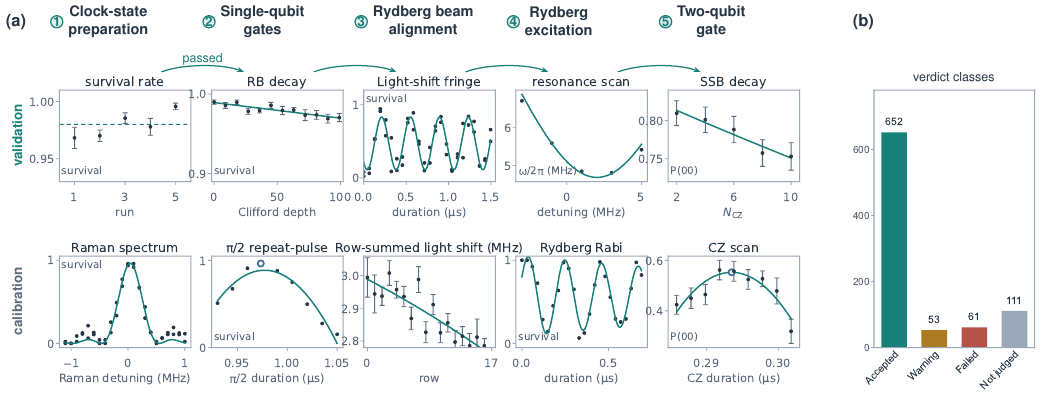}
\caption{\textbf{Representative results from the five experimental calibration Skills and the verdicts recorded over the deployment.} (a) The columns correspond to clock-state preparation, single-qubit gates, Rydberg beam alignment, Rydberg excitation, and the two-qubit gate. The upper validation row shows the state-preparation survival rate, randomized-benchmarking decay, light-shift Ramsey fringe, resonance scan, and symmetric-stabilizer decay; the lower calibration row shows the Raman spectrum, $\pi/2$ repeat-pulse curve, row-summed light shift, Rydberg Rabi oscillation, and CZ-duration scan. Points are measured values and curves are fits; error bars, where shown, are $1\sigma$ statistical uncertainties (Supplemental Material). The dashed line in the survival panel marks the predefined acceptance level. In the illustrated maintenance sequence, the next module is invoked only after the current validation satisfies its acceptance criterion. (b) Verdict classes of the 877 engagement records logged over the deployment, grouped as accepted, warning or inconclusive, failed, and not judged (definitions in the Supplemental Material).}
\label{fig:results}
\end{figure*}

We evaluate the system on a $^{87}$Rb neutral-atom array apparatus built around 852-nm optical tweezers, programmable arrays of more than 1000 sites, SLM-based rearrangement, Raman ground-state control at 795~nm, and two-photon Rydberg excitation through 420-nm and 1013-nm beams; the apparatus is shown in Fig.~\ref{fig:apparatus}, and full apparatus parameters are collected in the Supplemental Material.
We define a \emph{maintenance round} as one execution of the complete routine chain, issued as a single instruction: clock-state preparation, Rydberg beam alignment, and Rydberg excitation calibration.
Within this workflow, single-qubit randomized benchmarking provides gate-level validation, while the two-qubit module combines CZ-gate optimization with symmetric stabilizer benchmarking.
The \emph{deployment period} spans 90 days, over which the agent completed more than 850 experimental engagements; we count one engagement as a single execution by the agent, either a complete maintenance round or an individual calibration, verification, or readout task, and each instruction from the experimentalist may initiate one or several engagements.
Figure~\ref{fig:results}(a) collects representative results from the five experimental calibration Skills; Fig.~\ref{fig:results}(b) summarizes the verdict recorded for each of these engagements.

The calibration sequence begins with a preflight verification of the apparatus, corresponding to the preflight checks shown in Fig.~\ref{fig:system}: the agent checks that all instrument interfaces are reachable, reads the wavemeter channels, and confirms the MOT by absorption imaging and tweezer loading by fluorescence imaging.
The fluorescence images also provide the occupancy map used for site selection in the subsequent calibration and validation measurements.

Clock-state preparation follows the Raman transfer and Raman-assisted optical-pumping scheme of Refs.~\cite{levine2019,evered2023} (sequence in the Supplemental Material); the agent determines the transition frequencies and $\pi$-pulse durations of the transfer and auxiliary pulses.
Separately, a repeat-pulse procedure calibrates the $\pi/2$-pulse duration used for single-qubit control.
The first two columns of Fig.~\ref{fig:results}(a) show these calibration measurements in the lower row and their corresponding validation measurements in the upper row: state-preparation survival is compared with the predefined acceptance level, and single-qubit gate performance is evaluated by randomized benchmarking.

The calibration sequence then proceeds to the Rydberg lasers, beginning with alignment of the Rydberg beams to the array.
The agent measures the spatial light-shift profile of each Rydberg excitation beam by inserting a short pulse from the corresponding laser into a Ramsey sequence, and evaluates the alignment from the row- and column-resolved light-shift frequencies, commands a bounded mirror step, verifies the actuator read-back, and repeats the measurement until the spatial offsets satisfy the convergence criterion.

\begin{figure*}[t]
\centering
\includegraphics[width=\textwidth]{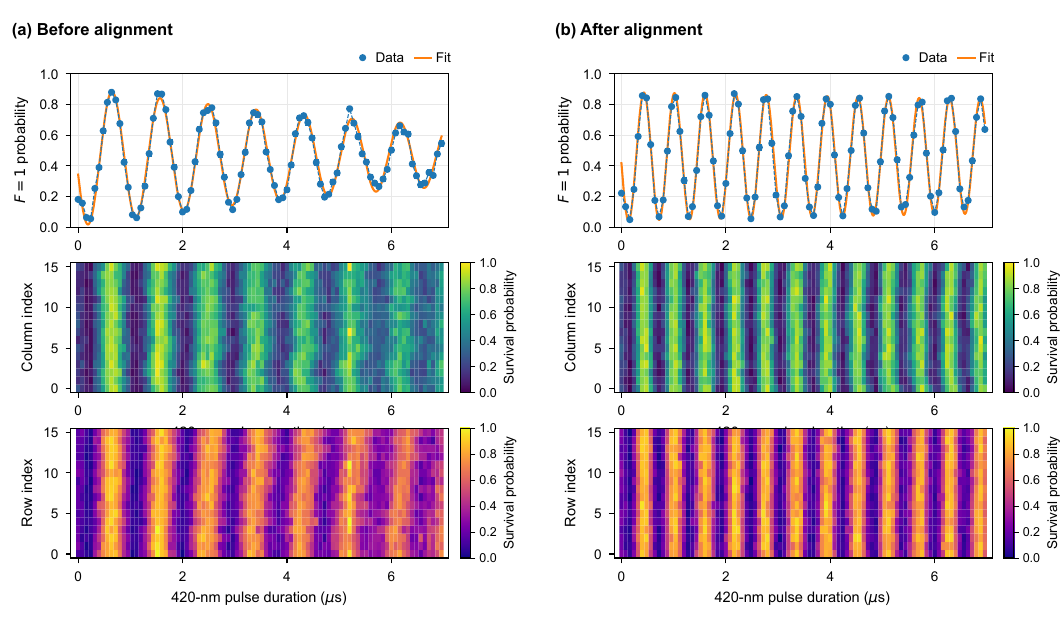}
\caption{\textbf{Spatial light-shift measurements for alignment of the 420-nm Rydberg beam.} Panels (a) and (b) show representative measurements before and after alignment, respectively. Each panel shows, from top to bottom, the spatially merged light-shift Ramsey oscillation, the column-resolved population map, and the row-resolved population map, as functions of the inserted 420-nm pulse duration. Blue points with error bars and orange curves in the top row show the measured populations, uncertainties, and fits, respectively, using the $F=1$ readout convention; the heat maps show the same $F=1$ probability on a common 0--1 color scale.}
\label{fig:beam_alignment}
\end{figure*}

The third column of Fig.~\ref{fig:results}(a) shows a representative light-shift Ramsey fringe and the row-summed light shift used for beam diagnostics.
Figure~\ref{fig:beam_alignment}(a) and (b) show representative spatially resolved light-shift measurements of the 420-nm beam before and after alignment, respectively, from which the agent judges whether the shift is uniform across the array.
After alignment, the agent characterizes the Rydberg excitation resonance and Rabi frequency: it fits a Rabi oscillation at each detuning and reads the on-resonance $\Omega_0$ and the offset $\delta_0$ from the resulting frequency-versus-detuning curve.
The fourth column of Fig.~\ref{fig:results}(a) shows representative resonance-scan and Rabi data.

Finally, the two-qubit module targets a time-optimal CZ gate evaluated by symmetric stabilizer benchmarking (SSB) following Ref.~\cite{tsai2025ssb} (see the Supplemental Material for the protocol).
The module performs a complete optimization round over six gate parameters, namely the amplitude of the time-optimal pulse, the Rabi scale, the detuning ratio, the initial phase, the CZ virtual phase, and the CZ duration, at fixed entangling order.
The parameters are scanned sequentially, and the atomic response acquired at each candidate configuration is evaluated using SSB.
The fifth column of Fig.~\ref{fig:results}(a) shows a representative CZ-duration scan and the SSB decay used for validation.
The configuration yielding the best SSB result was identified and reported.

We have demonstrated a language-agent system that executes recurrent maintenance of a neutral-atom quantum processor through reusable Skills, deterministic runners, and explicit safety and authorization constraints.
During a 90-day deployment, the agent completed more than 850 experimental engagements, spanning complete maintenance rounds and individual calibration, verification, and readout tasks.
Together with the representative clock-state, Rydberg, and gate-level results, these observations establish sustained and auditable agent-based operation across the local apparatus.
This extends agent-based quantum laboratory operation from one-time device bring-up~\cite{lzy_vibe} and single-step natural-language experiments~\cite{dalyac2026agentic} to recurrent maintenance of an optically controlled quantum processor.

The automation of calibration and maintenance provides an operational foundation for subsequent experiments and for maintaining the control quality required by larger-scale quantum workloads.
The present deployment evaluates operational feasibility and representative experimental outcomes rather than providing a controlled comparison of decision quality, human effort, or throughput against conventional operation.
With additional experiment-specific Skills and human supervision, the same architecture could support state-aware experimental assistance and more complex experimental campaigns; coordination among subsystem-specific agents is a further direction as the apparatus grows.

This work transfers the Skill paradigm from superconducting quantum processors~\cite{lzy_vibe} to a neutral-atom platform.
The modular design is intended to reduce reconstruction by localizing device-specific interfaces in coordination Skills while retaining reusable experimental procedures where applicable.
Trapped-ion systems, photonic platforms, and neutral-atom arrays share heterogeneous optical instruments, drift-dominated calibration cycles, and safety considerations for high-power lasers.
These shared characteristics suggest that the architecture may generalize into a common control layer for atomic, molecular, and optical (AMO) quantum computing platforms, although platform-specific procedures and acceptance criteria would require separate validation.

As laboratory agents move from isolated demonstrations to sustained operation, standardized evaluation must address decision quality, safety compliance, human intervention, and experimental throughput.
Existing efforts such as QCalEval~\cite{QCalEval} address calibration-plot interpretation; extending evaluation to end-to-end calibration and maintenance workflows remains an open challenge, and example tasks drawn from the present deployment are given in the Supplemental Material.
Such evaluations, together with auditable long-term deployments, will be essential for assessing the reliability and scalability of laboratory agents.

\begin{acknowledgments}
We thank Xinyang Wen, Haochen Wang, Ye Yang, and Yingdong Chen for their insightful discussions. This work is supported by the National Natural Science Foundation of China (NSFC; grant nos. 12325501, 92565306, 92476205, 92265205, and 92476001), the Strategic Priority Research Program of the Chinese Academy of Sciences, Grant No.XDB1680000, the National Key Research and Development Program of China (grant no. 2025YFA1411600), and the Quantum Science and Technology-National Science and Technology Major Project (grant no. 2021ZD0302104).
\end{acknowledgments}

\bibliographystyle{apsrev4-2}
\bibliography{refs}

\clearpage
\onecolumngrid
\setcounter{section}{0}
\setcounter{figure}{0}
\setcounter{equation}{0}
\renewcommand{\thesection}{S\arabic{section}}
\renewcommand{\thefigure}{S\arabic{figure}}
\renewcommand{\theequation}{S\arabic{equation}}
\renewcommand{\theHsection}{S\arabic{section}}
\renewcommand{\theHfigure}{S\arabic{figure}}
\renewcommand{\theHequation}{S\arabic{equation}}

\section*{Supplemental Material}
\addcontentsline{toc}{section}{Supplemental Material}
\section{Experimental platform details}
\label{app:apparatus}

The main 852-nm optical-tweezer system supports random-loading arrays of more than 1000 sites, at loading and cooling trap depths of 20--30~MHz (a representative calibration gives approximately 27.5~MHz); the tweezers are switched off during coherent Rydberg excitation.
The high-NA objective has NA $=0.65$, a focal length of 7.7~mm, and a working distance of 14~mm.
The measured base pressure is approximately $1.3\times10^{-11}$~mbar and the single-atom lifetime approximately 31~s.
A 20-ms fluorescence exposure yields 99.99\% imaging fidelity with 99.85\% survival in platform characterization, and a representative rearrangement converts a randomly loaded reservoir into a $16\times16$ target array at 98.9\% filling.

For ground-state control, the 795-nm Raman beam has a representative line-shaped profile at the atoms of approximately $75\times1622~\mu$m, a one-photon detuning of approximately $+180$~GHz, and circular incident polarization; the two frequency components are separated by approximately 6.835~GHz.
The bias field is approximately 7~G along the transverse east--west direction for the $\sigma^{\pm}$ transfer steps and is rotated to approximately 10~G along the south--north optical axis for the final clock-state transfer and Rydberg excitation.
For Rydberg excitation, the 420-nm beam (frequency-doubled from the 840-nm fundamental) has a representative elliptical waist of approximately $78\times250~\mu$m with $\sigma^{-}$ polarization, and the 1013-nm infrared branch is shaped as an approximately one-dimensional flat-top line of $281\times13~\mu$m with $\sigma^{+}$ polarization; the two beams counter-propagate along the row axis of the array.

All lasers are frequency-stabilized.
The 780-nm repump laser is locked to a saturated-absorption spectrum, and the 780-nm cooling laser is beat-note locked to the repump laser, so that its frequency follows the adjustable beat frequency.
The 795-nm optical-pumping laser is locked to a saturated-absorption spectrum and the 795-nm Raman laser to the wavemeter.
The 1013-nm laser and the 840-nm fundamental of the 420-nm laser are Pound--Drever--Hall locked to the same ultrastable cavity.
Because each laser is held at an absolute reference, every channel has a frequency at which the experiment expects it.
The wavemeter reads all channels together, and the Skill compares each reading with the stored lock point of that channel; a laser that has left its lock point invalidates every measurement taken after it, so this comparison is the first check the agent makes before an experiment.

The fluorescence image in Fig.~1 of the main text shows the rearranged $16\times16$ array, arranged in $4\times4$ blocks of $4\times4$ sites; it is a single shot. 
For display, a smooth background was subtracted and the intensity was scaled linearly, with the lowest 18\% of the display range set to background; no smoothing, dark-frame, or flat-field correction was applied.

\begin{figure}[htb!]
\centering
\includegraphics[width=\linewidth]{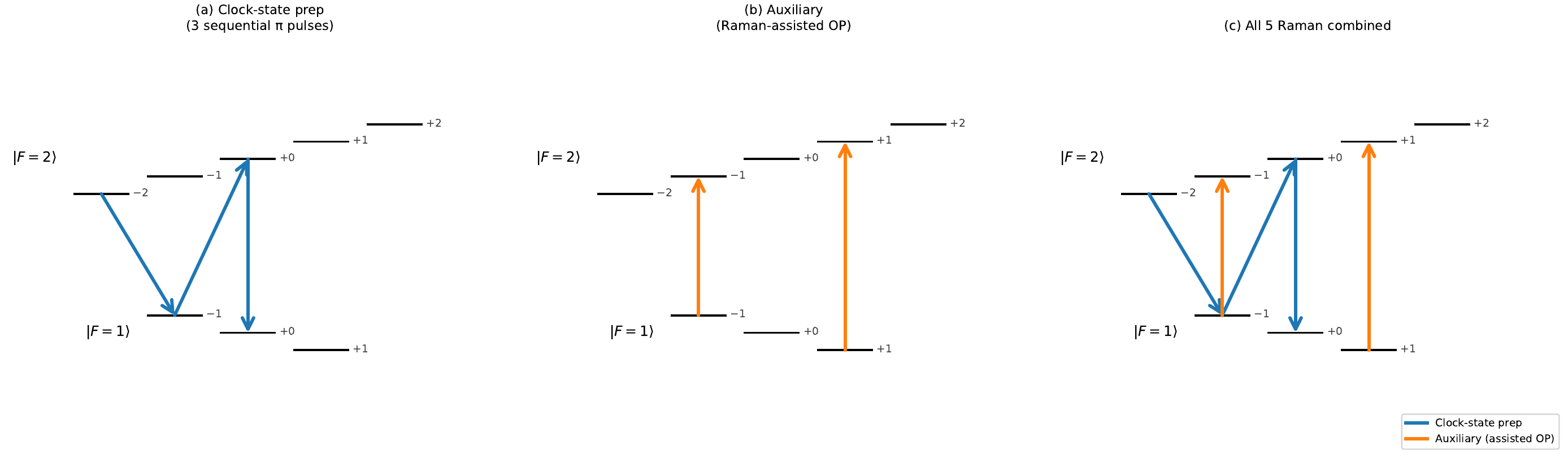}
\caption{\label{fig:raman_levels} Energy-level schematic of the five 795-nm two-photon Raman transitions acting on the $^{87}$Rb $5S_{1/2}$ ground state with Zeeman substructure.
(a)~Clock-state preparation: three sequential $\pi$ pulses (blue) transfer population along $|F=2,m_F=-2\rangle \rightarrow |F=1,m_F=-1\rangle \rightarrow |F=2,m_F=0\rangle \rightarrow |F=1,m_F=0\rangle$ ($\Delta m_F = +1,+1,0$).
(b)~Two auxiliary transitions (orange), $|F=1,m_F=\pm 1\rangle \rightarrow |F=2,m_F=\pm 1\rangle$ with $\Delta m_F = 0$, implement Raman-assisted optical pumping.
(c)~All five transitions combined.
Each arrow denotes one two-photon Raman transition through a virtual $5P_{1/2}$ level (not drawn); arrow slant encodes $\Delta m_F$, and the opposite tilts of the two hyperfine manifolds reflect the sign change of $g_F$.}
\end{figure}

Clock-state preparation follows the scheme of Refs.~\cite{levine2019,evered2023} (Fig.~\ref{fig:raman_levels}).
Atoms are optically pumped into $|F=2,m_F=-2\rangle$ and transferred by Raman $\pi$ pulses through $|1,-1\rangle$ to $|2,0\rangle$; after the quantization axis is rotated, a further Raman $\pi$ pulse transfers them to $|1,0\rangle$.
Residual population in $|1,\pm1\rangle$ is then removed by Raman-assisted optical pumping, in which two auxiliary Raman pulses drive $|1,\pm1\rangle\to|2,\pm1\rangle$, followed by resonant depumping from $F=2$.

\section{Hardware access and safety}
\label{app:hardware}

Instruments with standard remote protocols, such as the wavemeter and the oscilloscopes, are queried directly; instruments that run outside the experimental sequence under an existing control program, such as the SLM and the motorized mirror mounts, are reached through minimal command interfaces added to that program.
As an example of an instrument-specific limit, the mirror control program caps each piezo step of the beam-steering mirrors at 2~V; parameter and mirror writes are read back.
The wavemeter provides eight-channel frequency snapshots, which the agent compares with the known lock points.

\section{Labscript integration}
\label{app:labscript}

\begin{figure}[htb!]
\centering
\includegraphics[width=0.6\linewidth]{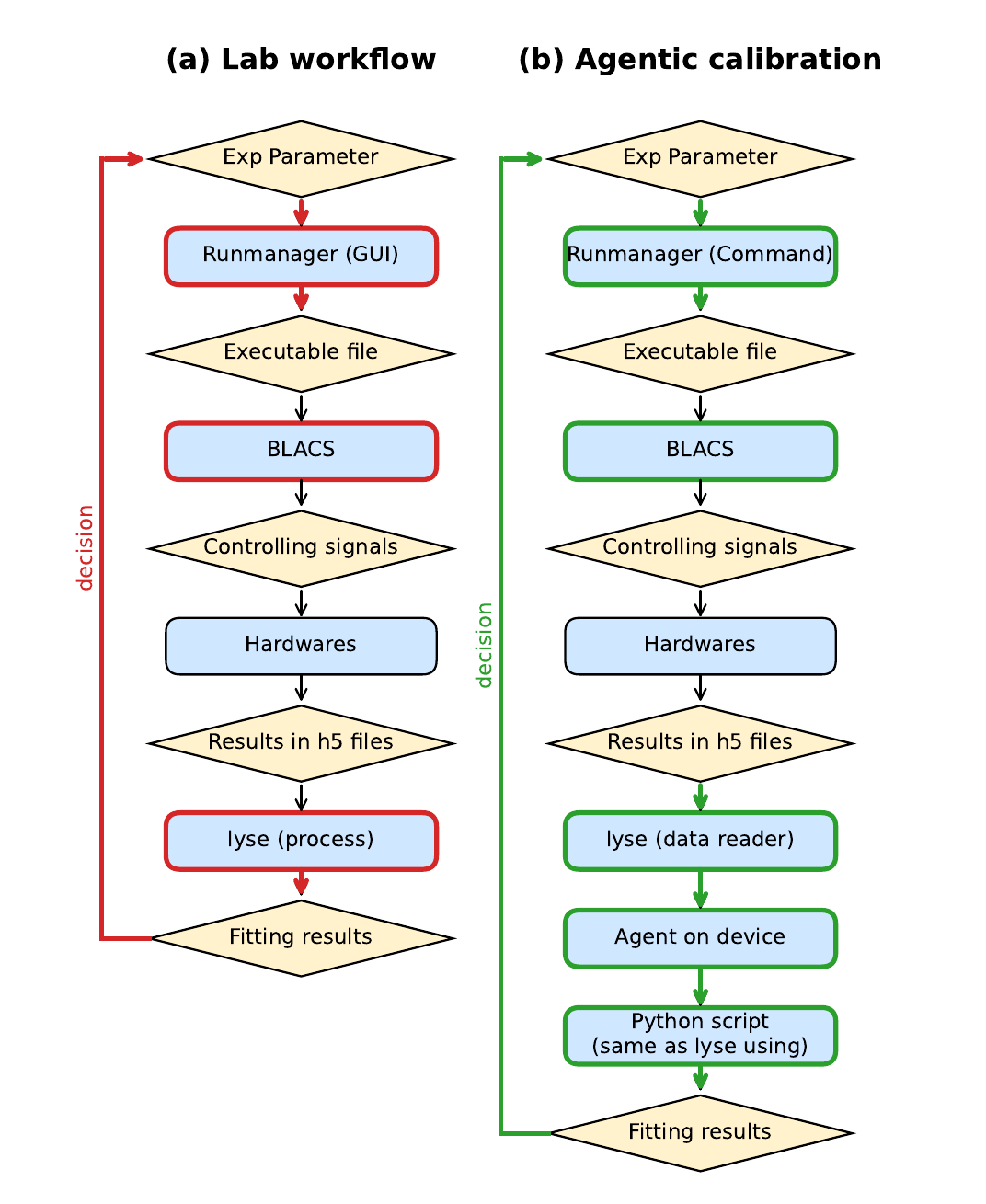}
\caption{\label{fig:workflow} \textbf{Manual (a) versus agentic (b) calibration workflow.} Runmanager compiles the experiment parameters into an executable, BLACS executes it hardware-timed, and the shot results are written to HDF5 files, with the fitted results feeding back into the next round of parameters. Color marks the operator of each stage, red for the human and green for the agent, from parameter setup and hardware engagement through data processing; in (b), the agent drives runmanager through its command interface and processes the shot files with the laboratory's existing analysis routines. In (b), the decision step is taken by the agent within documented procedures and otherwise referred to the experimentalist.}
\end{figure}

The agent interacts with the labscript control stack~\cite{labscript_suite} as a remote client through its Python interface.
The runmanager interface exposes labscript-file selection, global read and write with a raw flag, compilation and submission of shots, shot-folder retrieval, and the shot-count and error-state probes.
Per-shot analysis in the laboratory runs in the lyse analyzer, whose analysis routines are selected and managed through its graphical interface.
During the deployment, the agent carried out the same analysis by applying the laboratory's existing analysis routines to the HDF5 shot files (Fig.~\ref{fig:workflow}).
From the fitting results, the agent proposes the next round of parameters following the Skill.

\section{Agent configuration and logging}
\label{app:models}
\label{app:logging}
The agent ran in the Claude Code harness; daily operation and maintenance during the deployment used GLM-5.2, GLM-5.3, and GLM-5.3-flash.
The built-in memory of the harness was disabled, so that durable operating knowledge is maintained in the Skill documents rather than accumulated from past sessions.
For each executed run, the agent appends an entry to a daily Markdown log, which may include generated plots; the fields of each entry are specified in the Skill documents.

Figure~3(b) of the main text summarizes the 877 engagement records logged by the agent over the deployment.
Each verdict is recorded by the runner at a fixed decision point after the fit, and falls into one of four classes: accepted, when every registered check is satisfied; warning or inconclusive, when a value is returned but misses a criterion or cannot be judged; failed, when no acceptable value is obtained; and not judged, for entries that record no verdict.

\section{Gate benchmarking protocols}
\label{app:fits}

\subsection{Single-qubit gates: Clifford randomized benchmarking}

Single-qubit gate performance is evaluated using standard depth-dependent Clifford randomized benchmarking~\cite{magesan2011}.
Random gates are sampled from the 24-element single-qubit Clifford group and followed by a recovery Clifford that ideally returns the qubit to its prepared state.
Cliffords are compiled into Raman rotations about the $x$ and $y$ axes; the $\pi/2$ pulse is calibrated independently by a repeat-pulse procedure.
The sequences are generated offline and stored as a fixed library (20 random sequences of up to 200 Cliffords, seed 7); each shot applies the first $l$ gates of one sequence followed by the precomputed recovery Clifford, the group element that inverts the composite of the first $l$ gates.
The dataset in Fig.~3 of the main text uses depths $l \in \{0, 9, \ldots, 99\}$ with one shot per sequence and depth, each shot measuring all sites of the loaded array.
The per-shot survival is the fraction of initially occupied sites that remain occupied, so initially empty sites are excluded from the denominator while atom loss during the sequence counts as failure.
The site-pooled, depth-averaged survival is fitted by unweighted least squares to the depolarizing model
\begin{equation}
 P_1(l)=\frac{1}{2}+\frac{1}{2}\,(1-d_{\mathrm{SPAM}})(1-d)^l,
 \label{eq:rb_dep}
\end{equation}
with $d$ and $d_{\mathrm{SPAM}}$ free.
This is the standard RB decay $A p^l + B$~\cite{magesan2011} with the asymptote fixed at $B=1/2$, $p=1-d$, and $A=(1-d_{\mathrm{SPAM}})/2$; under the assumptions of the RB model, the fidelity per Clifford is $(1+p)/2 = 1-d/2$.
This is a per-Clifford quantity; conversion to a native-pulse fidelity requires additional compilation and noise assumptions.
Error bars are the standard error of the depth-averaged survival; per-column and per-row refits of the same model serve only as spatial-uniformity diagnostics.

\subsection{Two-qubit gates: symmetric stabilizer benchmarking}
The two-qubit benchmark is analyzed by the experiment's canonical binomial maximum-likelihood procedure with Fisher uncertainties, following the symmetric stabilizer benchmarking protocol of Ref.~\cite{tsai2025ssb}.
Return probabilities are measured at several fixed entangling-gate numbers $N_{\mathrm{CZ}}$ and fitted with
\begin{equation}
 P(N_{\mathrm{CZ}})=a_0 F^{N_{\mathrm{CZ}}},
 \label{eq:ssb_decay}
\end{equation}
with per-column and global results fitted separately.
$F$ denotes the fitted SSB decay parameter; its use as an estimate of CZ fidelity over symmetric input states follows the protocol-specific assumptions and is not automatically the average fidelity over the full two-qubit Hilbert space.
The circuit is randomized over the symmetric subspace and returns to the initial state $\ket{00}$, which is also the measured observable (return probability).
Atom pairs are formed along the array columns from the post-rearrangement image; only initially loaded pairs enter the count $n$, and pairs that lose an atom or leave the target state during the circuit remain in $n$ but out of the success count $k$, so atom loss and leakage are absorbed into the offset $a_0$ rather than post-selected.
At each $N_{\mathrm{CZ}}$ the likelihood is binomial in the pooled pair counts, with $p = a_0 F^{N_{\mathrm{CZ}}}$ and both $a_0$ and $F$ fitted by maximum likelihood in log space (multiple starts; points with $N_{\mathrm{CZ}} < 2$ excluded).
Uncertainties follow from the Fisher covariance (numerical Hessian of the negative log-likelihood with the delta method), and the reduced $\chi^2$ of the fitted curve serves as a consistency check.

\section{Rydberg beam alignment: spatially resolved light-shift Ramsey scans}
\label{app:ls_alignment}

The alignment diagnostic of Fig.~4 of the main text is a light-shift Ramsey sequence: two Raman $\pi/2$ pulses enclose an inserted 420-nm pulse of variable duration $t$, whose light shift $\delta$ advances the fringe phase by $2\pi\delta t$.
The Raman drive setpoint and the pulse sequence are held fixed while the beam position is changed, so the fringe frequency reports the light shift acting on the atoms.
Each scan spans $0\le t\le 6.96~\mu$s in 88 points with four shots per point, and the analyzed observable is the $F=1$ survival probability.
The site-merged curve is fitted by the damped sine
\begin{equation}
 P_1(t)=C+\frac{A}{2}\left[1-e^{-t/T_2}\cos(2\pi f t+\phi)\right],
 \label{eq:ls_ramsey}
\end{equation}
with all five parameters free, where $T_2$ is the $1/e$ envelope time and $f$ is identified with the light shift expressed as a frequency.
The fit weights each point by its binomial standard error $\sqrt{p(1-p)/n}$, with $n$ the number of occupied reference sites entering that point.
The column- and row-resolved maps are produced by averaging the survival over the orthogonal direction; their per-row and per-column fits use the same model without weighting and serve as uniformity diagnostics. We quantify the row-to-row variation as the range of $f$ over the 16 rows divided by its mean.

A point can be over-weighted by its own defect: when the pre-sequence reference image (\texttt{array\_map\_1}) registers far fewer atoms than the shot actually loads, the site-occupancy ratio after and before the sequence is artificially high and can exceed unity, with a correspondingly small binomial error.
Shots whose total \texttt{array\_map\_1} occupancy is below 60\% of the median of their own run are therefore discarded before averaging; each run is thresholded against its own median, which also allows runs with different repetition counts to be pooled.
This removes 15 of 352 shots from the before-alignment dataset and 4 of 352 from the after-alignment dataset.
The fitted fringe frequency changes by less than 0.05\% in either case, while for the after-alignment dataset $T_2$ increases from 50.2 to 58.1~$\mu$s.

\section{Example evaluation tasks}
\label{app:benchmark}

An agent operating a laboratory differs from a language model in conversation in that it invokes tools and acts on the returned results.
Evaluating such an agent therefore requires tasks that test not only knowledge and reasoning but also the selection and execution of the correct tool and the interpretation of its output.
Existing efforts such as QCalEval~\cite{QCalEval} evaluate vision-language models on the interpretation of calibration plots.
Drawing on the experience of the present deployment, we propose two benchmark dimensions that complement result interpretation for full-laboratory agents: instrument status interpretation and experiment operation design, both of which arise naturally when an agent is connected to heterogeneous instruments and entrusted with experiment execution.
Three illustrative tasks are presented below: one that extends result interpretation with a committed action, and one for each of the two dimensions.

\textbf{Result analysis and action decision.}
This task extends calibration-plot interpretation by requiring the agent to invoke a fitting tool and then commit to a downstream action.

\begin{taskbox}
\textbf{Materials:} recorded per-shot survival data from a one-dimensional Rydberg Rabi scan (75 shots, full array); the corresponding timing script; the Rydberg-excitation Skill document.

\medskip
\textbf{Question.} A recent calibration gives a reference value of $\Omega/2\pi \approx 3.5$~MHz.

\textbf{Q1.} Invoke the fitting tool and report $\Omega/2\pi$, the implied $\pi$-pulse duration, and a measure of the fit quality.

\textbf{Q2.} Given the fitted values and the reference value, select one of three actions: accept the fit, repeat the scan, or investigate the hardware. Justify in two sentences.

\medskip
\textbf{Reference answer.}
The fit gives $\Omega/2\pi$ within a few percent of the reference value, a $\pi$-pulse duration of order $0.1~\mu$s, and a moderate fit quality, with no systematic structure in the residuals.

\emph{Must contain:} the fitted values and their agreement with the reference value; the decision to accept the fit; and the supporting reasons, namely that the survival at $t=0$ is close to unity and the oscillation is resolved, which rules out a hardware fault, and that the moderate fit quality follows from the small number of repetitions per point.
\end{taskbox}

\emph{Deployment outcome.} The agent fitted the scan, proposed acceptance, and the experimentalist reviewed and approved the decision. With a script written ad hoc, a single failed fit was sometimes reported as a failed experiment; after the agent was required to use the laboratory's existing routine, which retries the fit with different seeds, its decisions stabilized.

\textbf{Instrument status interpretation.}
This task tests whether the agent can read a status snapshot of an instrument, classify each channel, and recognize where the data stop being sufficient.
What it lacks here is not knowledge of the instrument, which the Skill records, but knowledge of the present state of the laboratory.

\begin{taskbox}
\textbf{Materials:} two recorded wavemeter snapshots taken 7~s apart; the channel-to-laser mapping, the lock points, and the per-channel tolerances recorded in the wavemeter Skill.

\medskip
\textbf{Question.} Six of the eight wavemeter channels are connected; the near-lock and warning tolerances are $\pm 10$~MHz and $\pm 30$~MHz for most channels.

\textbf{Q1.} For each connected channel, report the absolute frequency, the offset from its lock point, and its class.

\textbf{Q2.} State whether the two snapshots determine the cause of any anomaly.

\textbf{Q3.} State the next action.

\medskip
\textbf{Reference answer.}
Five channels lie within the near-lock tolerance in both snapshots.
The cooling laser lies about $0.11$~GHz from its lock point in the first snapshot and about $0.57$~GHz in the second, far outside the warning tolerance, and the second reading is further from the lock point than the first.

\medskip
{\footnotesize
\setlength{\tabcolsep}{3pt}%
\begin{tabular}{l r r r r r l}
\hline
 & Lock point & \multicolumn{2}{c}{First snapshot} & \multicolumn{2}{c}{Second snapshot} & \\
\cline{3-6}
Laser & (THz) & $f$ (THz) & $\delta f$ (MHz) & $f$ (THz) & $\delta f$ (MHz) & Class \\
\hline
780 cooling & 384.227936 & 384.2278287 & $-107$ & 384.2273707 & $-565$ & abnormal \\
780 repump & 384.234605 & 384.2346017 & $-3.3$ & 384.2346019 & $-3.1$ & normal \\
795 Raman & 377.285050 & 377.2850502 & $+0.19$ & 377.2850498 & $-0.24$ & normal \\
795 OP & 377.105041 & 377.1050377 & $-3.3$ & 377.1050380 & $-3.0$ & normal \\
840 ($\to$ 420) & 356.639073 & 356.6390766 & $+3.6$ & 356.6390765 & $+3.5$ & normal \\
1013 & 296.007990 & 296.0079896 & $-0.38$ & 296.0079899 & $-0.14$ & normal \\
\hline
\end{tabular}}

\medskip
The actual cause was a hand adjustment of the cooling laser's lock point, caught part-way: in the first snapshot the laser is still travelling, and the second reading sits at a second lock state about $565$~MHz away, which the same laser has taken on other occasions.
The other channels did not move, so nothing had failed.

\medskip
\emph{Must contain:} the cooling laser is outside its warning tolerance in both snapshots, at about $0.11$~GHz and $0.57$~GHz from its lock point, and the second reading is further from the lock point than the first; the other five channels are unchanged to within $0.5$~MHz between the two snapshots, so the anomaly is local to that channel; the snapshots do not determine the cause, because the agent has no record of what was done to this laser in the preceding minutes; the next action is to halt and ask the experimentalist whether the laser was adjusted, and to begin no experiment before the answer.

\emph{Not accepted:} declaring an instrument fault, concluding that the laser is settling, or starting an experiment on these data.
\end{taskbox}

\emph{Deployment outcome.} The agent took the first snapshot on its own and, without being asked, took the second 7~s later to see whether the reading was still moving; it classified the channels, could not determine the cause, and asked the experimentalist, who explained the situation and told it to continue.

\textbf{Experiment configuration.}
This task tests whether the agent understands the experimental sequence by its physical meaning rather than by parameter names, and whether it can identify the parameters of a planned experiment and detect residual settings left by the previous one.

\begin{taskbox}
\textbf{Materials:} the reference timing script for a coarse Rydberg resonance scan; a current globals configuration file; the Rydberg-excitation Skill.

\medskip
\textbf{Question.} The scan is to be run with the excitation pulse held at the $\pi$-pulse duration calibrated in the first task, with the 420-nm detuning stepped over a coarse range around the expected resonance, and with all other parameters as in the reference script.

\textbf{Q1.} Describe what the experimental sequence should be and which parameter should be scanned.

\textbf{Q2.} After reading the globals file, identify the name of the global variable that controls the scan axis.

\textbf{Q3.} Compare the loaded parameters with the target and list those that require modification, with their meaning, current value, and correct value.

\medskip
\textbf{Reference answer.}
\emph{Q1:} The sequence is full clock-state preparation (Raman $\pi$ chain with assisted optical pumping), followed by a single square two-photon Rydberg excitation pulse of fixed short duration with the 1013-nm laser overlapping, and survival-based readout (Rydberg atoms are lost, so survival dips at resonance).
The scan parameter is the 420-nm detuning.

\emph{Q2:} \texttt{detuning\_420\_for\_resonance\_MHz}.

\emph{Q3:} three parameters require correction, as listed below.

\medskip
\begin{tabular}{lll}
\hline
Parameter & Current & Correct \\
\hline
Excitation pulse duration & list of values & fixed at the $\pi$ pulse \\
Readout mode & state-selective & survival \\
AWG pre-trigger offset & present & removed \\
\hline
\end{tabular}

\medskip
The first silently introduces a second scan axis and raises the shot count by the length of the list; the second is a readout setting carried over from the previous experiment and is inconsistent with the survival readout that the scan requires; the third is an internal script constant that does not belong among the control-stack parameters.

\emph{Must contain:} the scan parameter, the three settings with their correct values, and the consequence of each.
\end{taskbox}

\emph{Deployment outcome.} This task was repeated many times in the deployment. Without the parameter meanings in the Skill, the agent took about several minutes and repeatedly stopped to ask the experimentalist which settings had to be changed; with the parameter meanings documented in the Skill, it identified the settings directly.

The development of new experimental scripts is equally relevant for a laboratory agent but depends strongly on the experimental environment, so it is not included here.

\section*{Data availability}
The data that support the findings of this study are available from the corresponding authors upon reasonable request.

\end{document}